\documentclass[nofootinbib,floatfix,twocolumn]{revtex4}

\usepackage{empheq} 
\usepackage{mathrsfs}
\usepackage{verbatim}
\usepackage{amssymb}
\usepackage{graphicx}
\usepackage{subfigure}
\usepackage{color}
\usepackage{rotating}

\usepackage[mathscr]{euscript}
\usepackage{amsmath}

\def\bsone{{\bf S_1}}
\def\bstwo{{\bf S_2}}

\def\be{\begin{equation}}
\def\ee{\end{equation}}
\def\ba{\begin{eqnarray}}
\def\ea{\end{eqnarray}}

\begin{document}
\bibliographystyle{apsrev}

\title{Zoom-Whirl Orbits in Black Hole Binaries}

\author{James Healy${}^{1}$, Janna Levin${}^{2,3}$, and Deirdre Shoemaker${}^{4}$}
\affiliation{${}^{1}$ Center for Gravitational Physics, The
  Pennsylvania State University, University Park, PA 16802}
\affiliation{${}^{2}$ Barnard College of Columbia University,
  Department of Physics and Astronomy, 3009 Broadway,
  New York, NY 10027}
\affiliation{${}^{3}$ Institute for Strings, Cosmology, and
  Astroparticle Physics,
Columbia University,
  New York, NY 10027}
\affiliation{${}^{4}$ Center for Relativistic Astrophysics and School
  of Physics, Georgia Institute of Technology, Atlanta, GA 30332}

\begin{abstract}

  Zoom-whirl behavior has the reputation of being a rare
  phenomenon. The concern has been that gravitational radiation would
  drain angular momentum so rapidly that generic orbits would
  circularize before zoom-whirl behavior could play out, and only rare
  highly tuned orbits would retain their imprint. Using full numerical
  relativity, we catch zoom-whirl behavior despite dissipation. The
  larger the mass ratio, the longer the pair can spend in orbit before
  merging and therefore the more zooms and whirls seen. Larger spins
  also enhance zoom-whirliness. An important implication is that these
  eccentric orbits can merge during a whirl phase, before enough
  angular momentum has been lost to truly circularize the orbit.
  Waveforms will be modulated by the harmonics of zoom-whirls, showing
  quiet phases during zooms and louder glitches during whirls.

\end{abstract}

\maketitle


Kepler's laws describe closed elliptical planetary motions.
A small relativistic correction accounts
for the tiny, anomalous precession of Mercury's perihelion.
If the sun is replaced by a black hole, the geodesic motions can
zoom and whirl in an extreme form of 
precession -- whirling around the center of mass in nearly
circular inspiral before zooming out along elliptical leaves. 
Zoom-whirl behavior
is characteristic 
of strong relativity and could potentially be detected in
the harmonics of the gravitational waves generated.

There are astrophysical settings that could populate eccentric
merges, such as dense galactic nuclei \cite{O'Leary:2008xt} or globular clusters \cite{Kocsis:2006hq,wen2003}. 
Consequently, it is an
important astrophysical question to ask: 
Can zoom-whirl behavior,
an intrinsically eccentric phenomenon,
survive the dissipative drain of gravitational radiation? 
In this article we report on results of numerical
relativity that show zoom-whirl behavior in comparable mass binaries,
answering this question in the affirmative.

As there is no analytic description of the curved
spacetime around two black holes, we rely either
on analytic approximations or on numerical
relativity to describe comparable black hole pairs.
Zoom-whirl behavior has been studied in extreme-mass ratio
inspirals~\cite{2002PhRvD..66d4002G,2004PhRvD..69h2005B,Burko:2006ua,Haas:2007kz}
and was recently found in an analytic approximation, specifically
conservative Post-Newtonian (PN)
approximations to black hole binaries
\cite{levin2008:2,grossman2008}.  Now we find
zoom-whirl orbits in full numerical relativity of spinning pairs.
Zoom-whirl behavior has already been found in numerical relativity for
equal-mass, nonspinning binaries in \cite{pretorius2007}. In that work, the initial
conditions were carefully tuned to find a special orbit, the
separatrix between bound orbital motion and plunge. The separatrix is
studied in detail in the PN approximation in \cite{grossman2008} and an analytic
solution for the separatrix in Kerr systems was found in
\cite{Levin:2008yp,PerezGiz:2008yq}.

In this Letter we show that zoom-whirl behavior in spinning pairs 
is a common feature of
eccentric orbits \cite{Levin:2008yp,levin2008:2,grossman2008}, despite
dissipation. In particular,
zoom-whirl orbits happen well away from the separatrices and so do
not in general require fine tuning of initial conditions (see also~\cite{sperhake-09,gold}).
Due to the computational expense of running
these simulations, a full scan of parameters is not possible. To focus
our investigation, we rely on analytic approximations
to
estimate good initial conditions
 and we then
run full numerical simulations to easily locate zoom-whirl inspirals. 
The further utility of the analytic
estimates is the transparency of interpretation. 

The anatomy of zoom-whirl behavior was quantified in Ref.\ \cite{levin2008}
 where it was shown that every orbit of a given $L$ can be described
 by one number that specifies the precession of the orbit per radial
 cycle from apastron to apastron. The amount by which an orbit will
 precess, that is, overshoot the previous apastron is
\begin{equation}
\Delta \phi_{precess}=2\pi q\quad {\rm where}\quad q=w+\frac{v}{z} \quad .
\end{equation}
A perfectly periodic orbit looks like a closed 1-leaf clover or 2-leaf
clover or 3-leaf clover, or $z$-leaf clover. 
And each
periodic orbit corresponds to a 
rational $q$ made up of $w$ integer number of nearly circular whirls
close to perihelion per leaf in the $z$-leaf clover. 
The vertex $v$ is more subtle and indicates the order in which the $z$-leaves are
traced out. So, a simple 3-leaf clover for instance is a $q=1/3$
($w=0,v=1,z=3$) and precesses past the previous apastra
by $\Delta \phi= 2\pi q=2\pi/3$ per radial cycle.
A 3-leaf clover that skips a
leaf in the pattern each time corresponds to $q=2/3$ ($w=0,v=2,z=3$)
and precesses by $\Delta \phi= 2\pi q=4\pi/3$ per radial cycle.
Therefore, $q$ quantifies
zoom-whirl behavior, the integer part signals the whirls per leaf and
the fractional part signals the number and order of the leaves. 

Another way to interpret the number is as the ratio of frequencies,
$q={\omega_\phi}/{\omega_r}$,
where $\omega_\phi$ is the average of the angular frequency per radial
cycle and $\omega_r=2\pi/T_r$
where $T_r$ is the time between apastra. For periodic orbits the
frequencies are rationally related and the orbit will eventually
close. Whirls accumulate near perihelion simply because the 
angular velocity is greatest on closest approach and the circumference
smallest.

Generic orbits are not periodic and do not
correspond to rational $q$. However, any generic orbit can be
approximated by a nearby periodic, just as any irrational number can be
approximated by a nearby rational number. Kepler's ellipse corresponds to
$q=0$ since it does not precess at all.  Mercury's precessing orbit
corresponds to $q\sim 10^{-7}$ since it precesses very little.
Technically, zoom-whirl behavior corresponds to $q>1$, so
there is at least
one whirl, although we will generally be interested in any substantial
precession, say $q>1/4$. 

{\it If} angular momentum is fixed,
$q$ decreases monotonically with decreasing 
energy. However angular momentum decreases, along with energy, as 
gravitational radiation is emitted and the evolution of $q$ will
not necessarily be monotonic. 
In the case of comparable
mass black hole pairs, $q$ will change quickly due to the rapid losses
to gravitational waves as evident in the simulations. 

The range of zoom-whirl behavior is most easily targeted with an
effective potential picture.
In Ref.\ \cite{levin2008,grossman2008} an effective potential method was used to describe
the center-of-mass motion of black hole binaries in the conservative 3PN
Hamiltonian description with spin-orbit coupling included.
There it was shown that at the turning points of the orbital motion,
the Hamiltonian itself is an effective potential
and the motion can be read off as easily as interpreting a ball
rolling along hills. 
\begin{figure}
  \centering
\includegraphics[width=50mm]{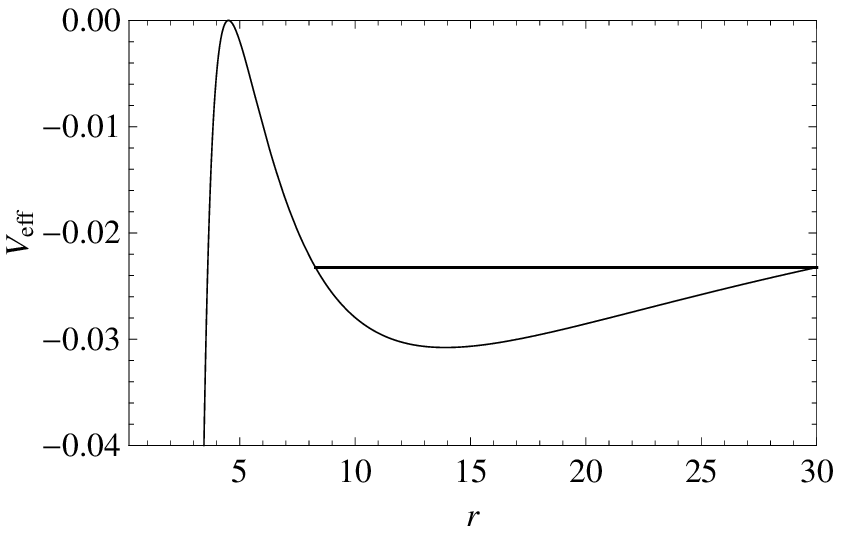}
\includegraphics[width=30mm]{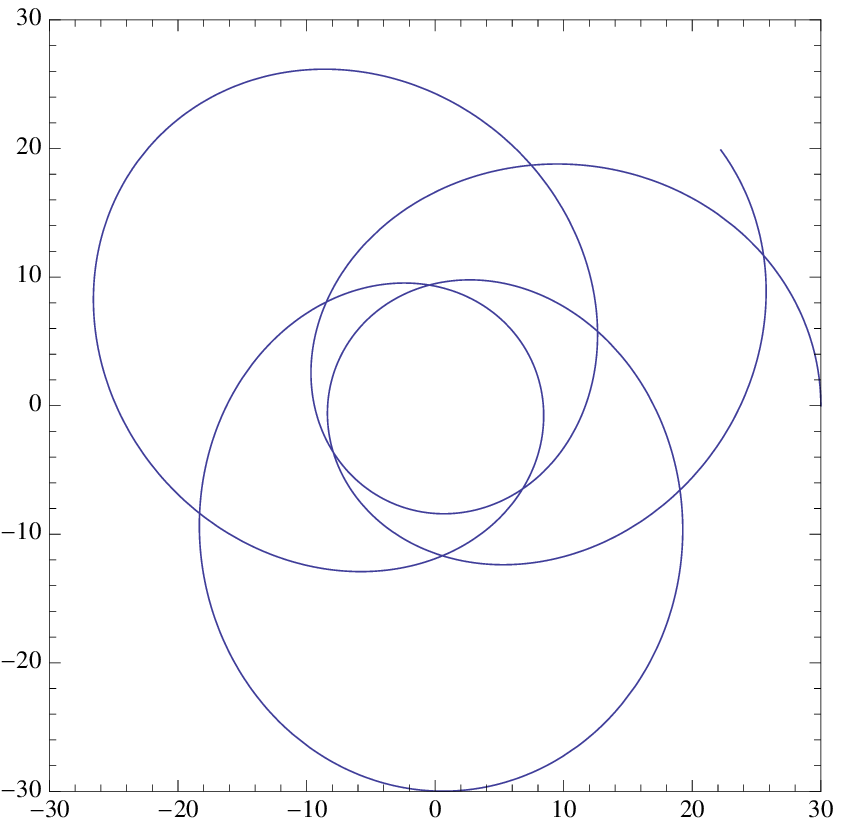}
\includegraphics[width=50mm]{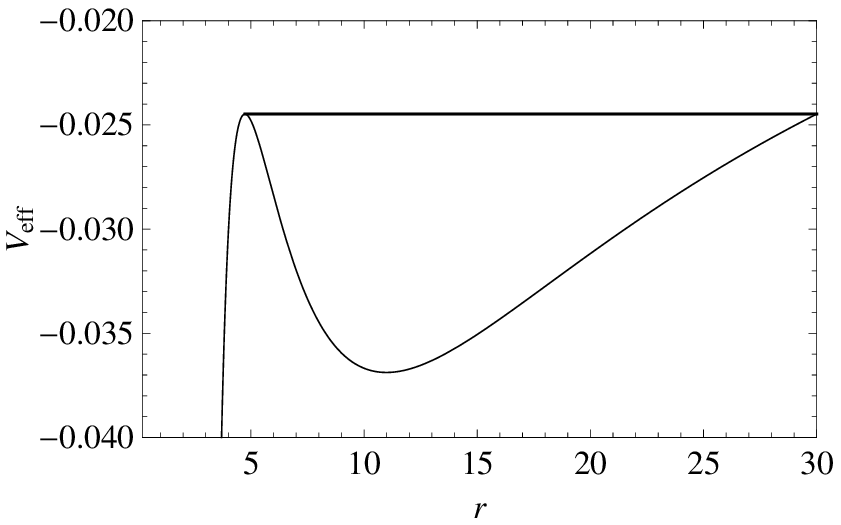}
\includegraphics[width=30mm]{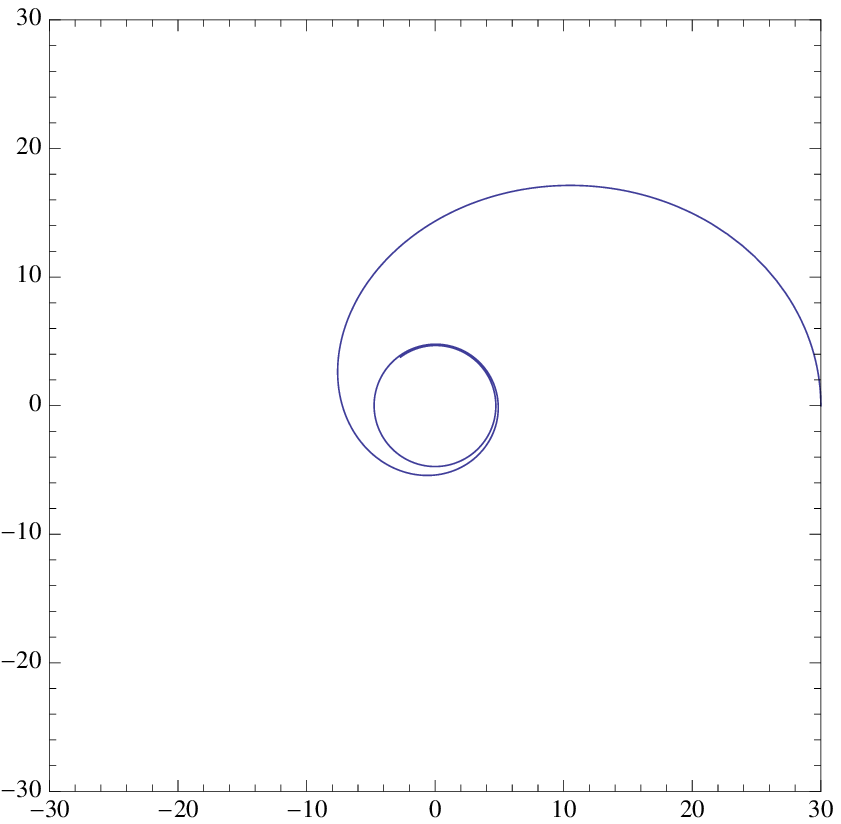}
\hfill
  \caption{The 3PN effective potential. 
Top:
$L_{IBCO}\sim 4.4$.
The straight line corresponds to the orbit $r_a=30$.
Bottom:
$L\sim 4.1$, for which the orbit with $r_a=30M$
is the separatrix.}
  \label{Vm.25Sm8}
\end{figure}

Consider the effective potential on the top left of
Fig.\ \ref{Vm.25Sm8} from the conservative 3PN Hamiltonian
\cite{buonanno2006,damoureob2001,levin2008,grossman2008}.
There is clearly a minimum of that potential which defines a stable
circular orbit.
An orbit energetically above the stable circle will execute elliptical
precessions as shown on the top right. Note these precessions are much
more extreme than Mercury's with 
$q> \frac{1}{3}$.
It looks like a precession around a three leaf
clover. 
The aperiodic orbit will eventually fill out an annulus.

Also evident is a maximum of the potential, an
unstable
circular orbit. 
For this angular momentum the unstable circular orbit is marginally, energetically
bound since $V_{\rm eff}$ just skims zero there, and has been called
an innermost
bound circular orbit (IBCO).
An orbit at rest at
infinity would asymptotically approach this circular orbit. And though
initially of eccentricity one, $e=1$, this orbit whirls an
infinite number of times as it climbs the potential toward the
unstable circle at the top. This separatrix is the infinite whirl
orbit, $q=\infty$, and is also known as a homoclinic orbit of eccentricity one. 

For angular momenta below this critical value, the unstable
circular orbit marches down in energy. For each unstable circle, there is a
corresponding separatrix of decreasing eccentricity.
As an example, on the bottom 
of Fig.\ \ref{Vm.25Sm8},
the separatrix
between bound motion and plunge has apastron $r_a=30M$ and
asymptotically approaches the unstable circular orbit in the infinite
whirl limit. So this is the $q=\infty$ orbit for $L\sim 4.1$, 
where throughout we measure angular momentum in units of $\mu M$ and
$M=m_1+m_2$ while $\mu=m_1 m_2/M$.
The homoclinic separatrix
is plotted alongside the corresponding
effective potential on the bottom right  
of Fig.\ \ref{Vm.25Sm8}.

We expect to see
zoom-whirl behavior until the unstable circular orbit and the stable
circular orbit merge at the ISCO (innermost stable circular orbit),
the separatrix with $e=0$.
Quasi-circular inspiral
plunges at the ISCO and so this inflection point in the
effective potential has received preferential attention. However,
orbits that maintain an
eccentricity during inspiral will merge by rolling over the top of the
potential, through a homoclinic separatrix, behavior we observe in our simulations.

Roughly then, for given external parameters
($m_2/m_1,\bsone,\bstwo$), zoom-whirl behavior should be sought in the
range 
$L_{ISCO}<L<L_{IBCO}$. One more initial condition needs to be
specified to define an orbit and that could be either the energy or
the apastron. We'll choose to fix the apastron.
We'll fill in the details for a fiducial example and then flush out
more general trends we have observed.

The procedure is the following: (1) choose external parameters
namely mass ratio and spins
($m_2/m_1,\bsone,\bstwo$), (2) render the effective potential
for the 3PN Hamiltonian with
spin-orbit couping, (3) glean the range of $L$ for which there will be
zoom-whirl behavior and (4) run simulations for this range of initial conditions.

For our fiducial example we take mass ratio $m_2/m_1=1/3$ and spin magnitudes
$S_1/m_1^2=S_2/m_2^2=0.3$ with spins antialigned with the
orbital angular momentum. Antialigned spins, like aligned spins, will
retain the orbital motion in the equatorial plane making it easier to
distinguish whirl precession. Other than the choice to retain
equatorial motion for transparency, there is nothing special about
the fiducial 
configuration.
We consider a set of orbits all of which begin at apastron $r_a=30M$.
Since
the simulations are so costly, we tighten the range to the
more realistic values of angular momenta below $L_{IBCO}$ and above the
value of the angular momentum at which $r=r_a$ is the apastron of a
homoclinic orbit. Specifically, we look between the values
represented by the top and bottom of Fig.\ \ref{Vm.25Sm8}; $4.1<L<4.4$.

There are a few reasons why the actual range for numerical relativity
will be offset from this 3PN prediction. For one,
dissipation ensures that $L$ changes as the orbit evolves. This is
equivalent to the effective potential dropping as the orbit
evolves. For another, the 3PN approximation is by definition not exact
and its predictions are not to be overstated in this strong-field
regime. Finally, although spin-orbit coupling is included in our
analysis, spin-spin coupling terms 
are not included since they introduce angular dependencies that 
spoil the effective potential
\cite{levin2008,grossman2008}.
Since spin-spin terms are small, their effect will presumably result in a
small perturbation to our 3PN effective potential.
Our range is intended only as a guide and apparently does
well enough, as we'll see.

In between this range, all orbits with our external parameters
$(m_2/m_1=1/3,S_1/m_1^2=S_1/m_2^2=0.3)$, antialigned, and an apastron of
$r_a=30M$, will show zoom-whirl behavior
with $1/3<q<\infty$. 
This is the story told by the conservative
dynamics. To test the survival of zoom-whirl orbits under the
dissipative effects of radiation reaction we turn to full numerical
relativity now.

\begin{figure}
 \centering
\includegraphics[width=30mm,angle=-90]{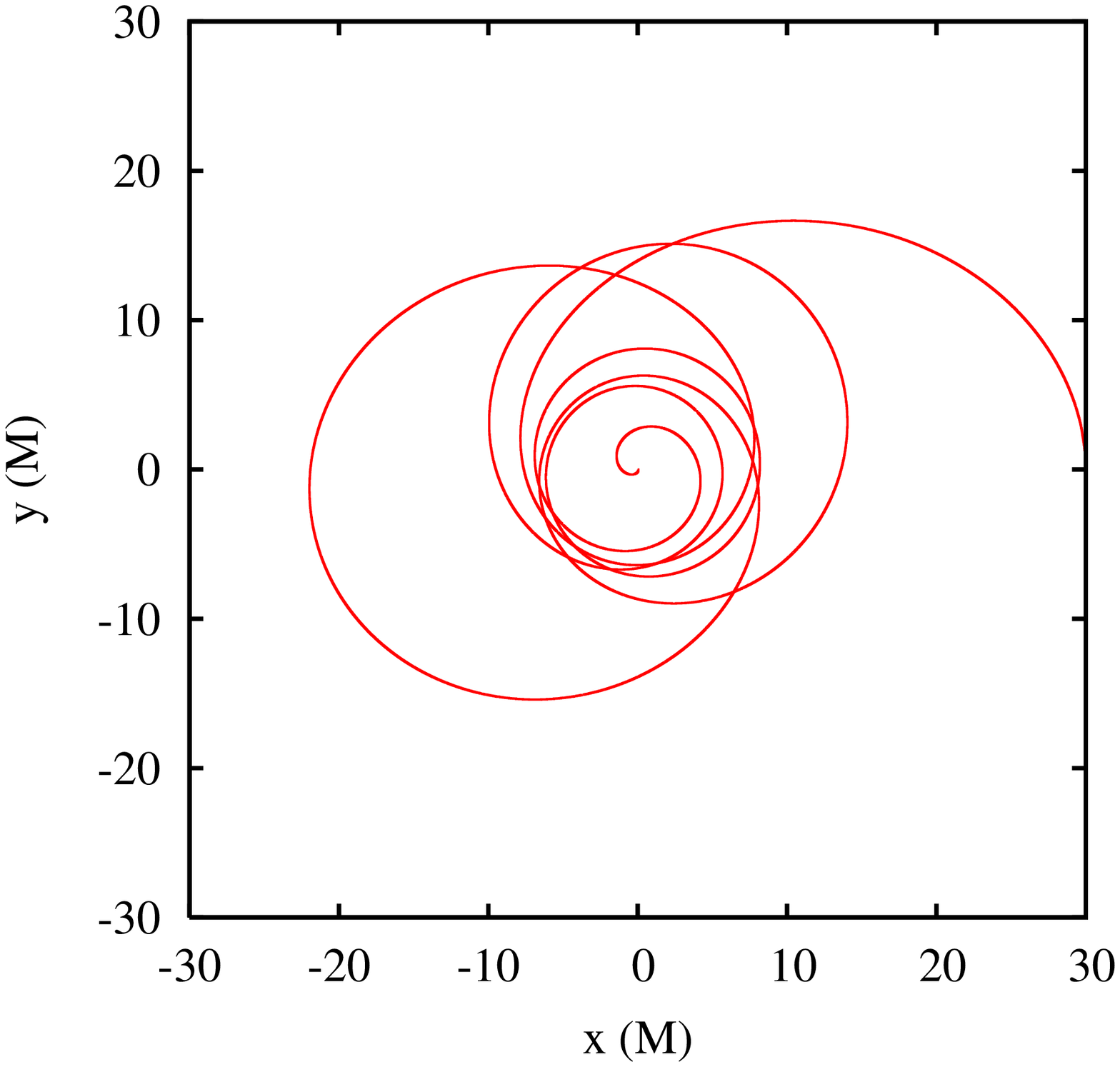}\includegraphics[width=30mm,angle=-90]{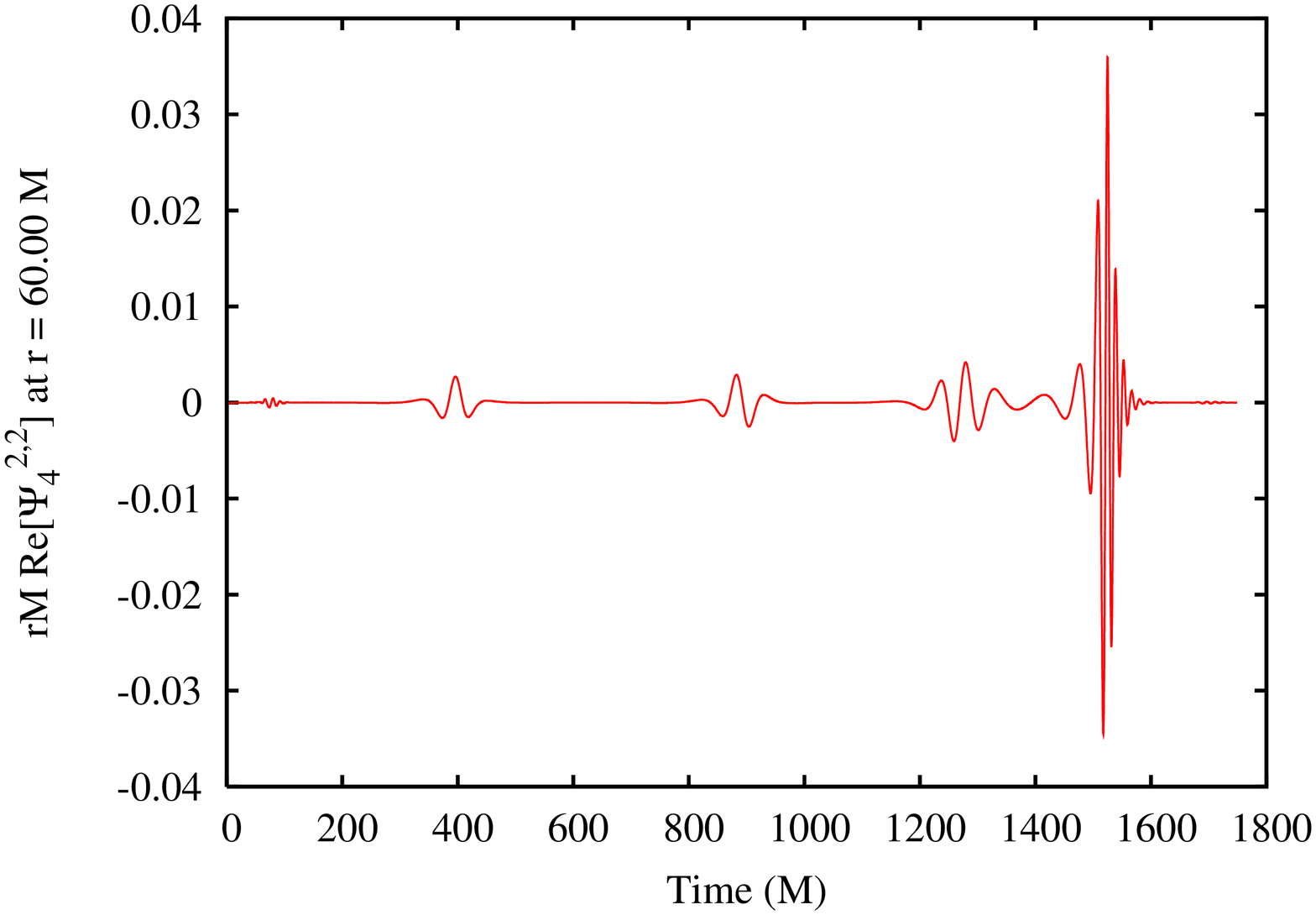}
\includegraphics[width=30mm,angle=-90]{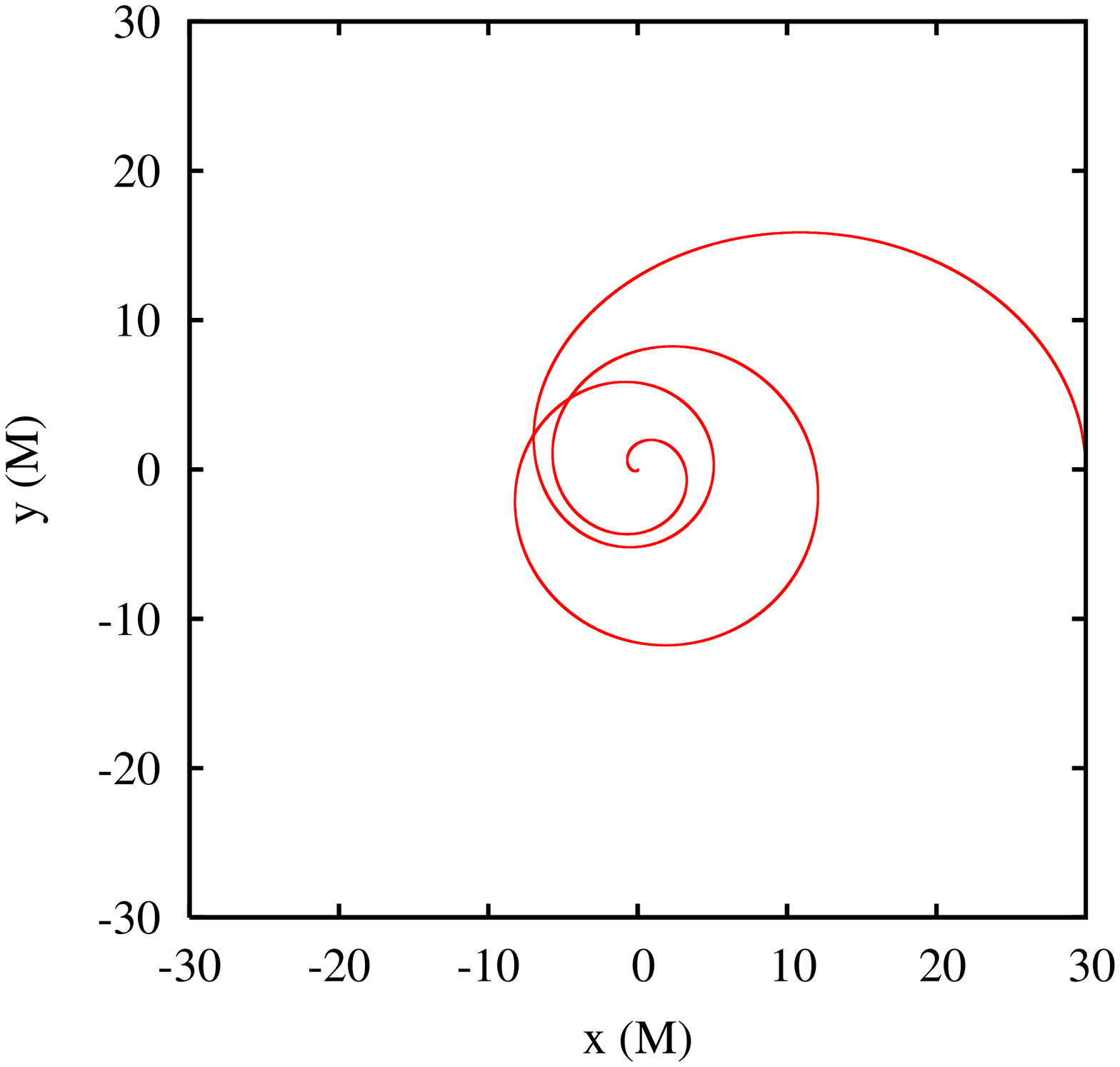}\includegraphics[width=30mm,angle=-90]{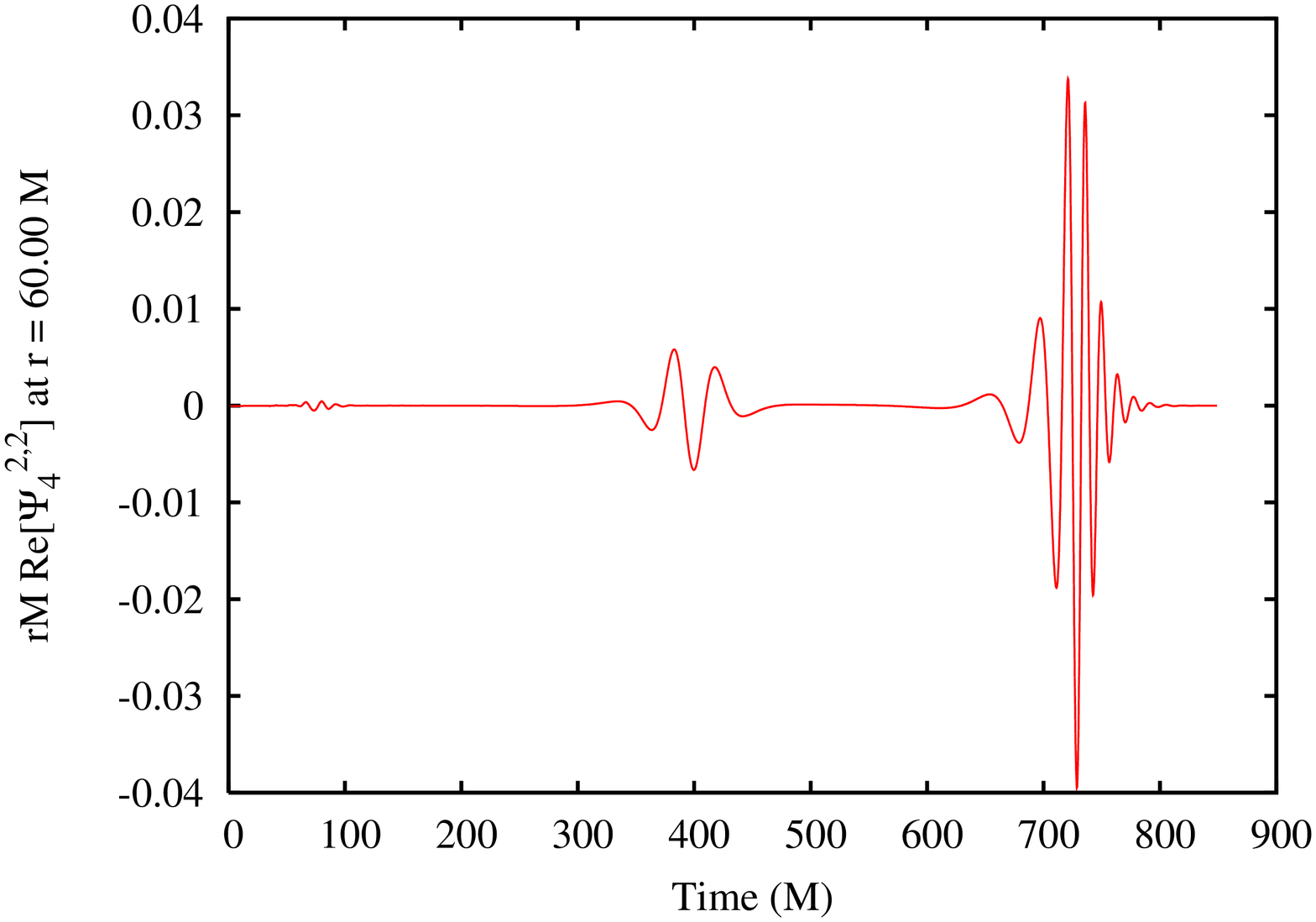}
\includegraphics[width=30mm,angle=-90]{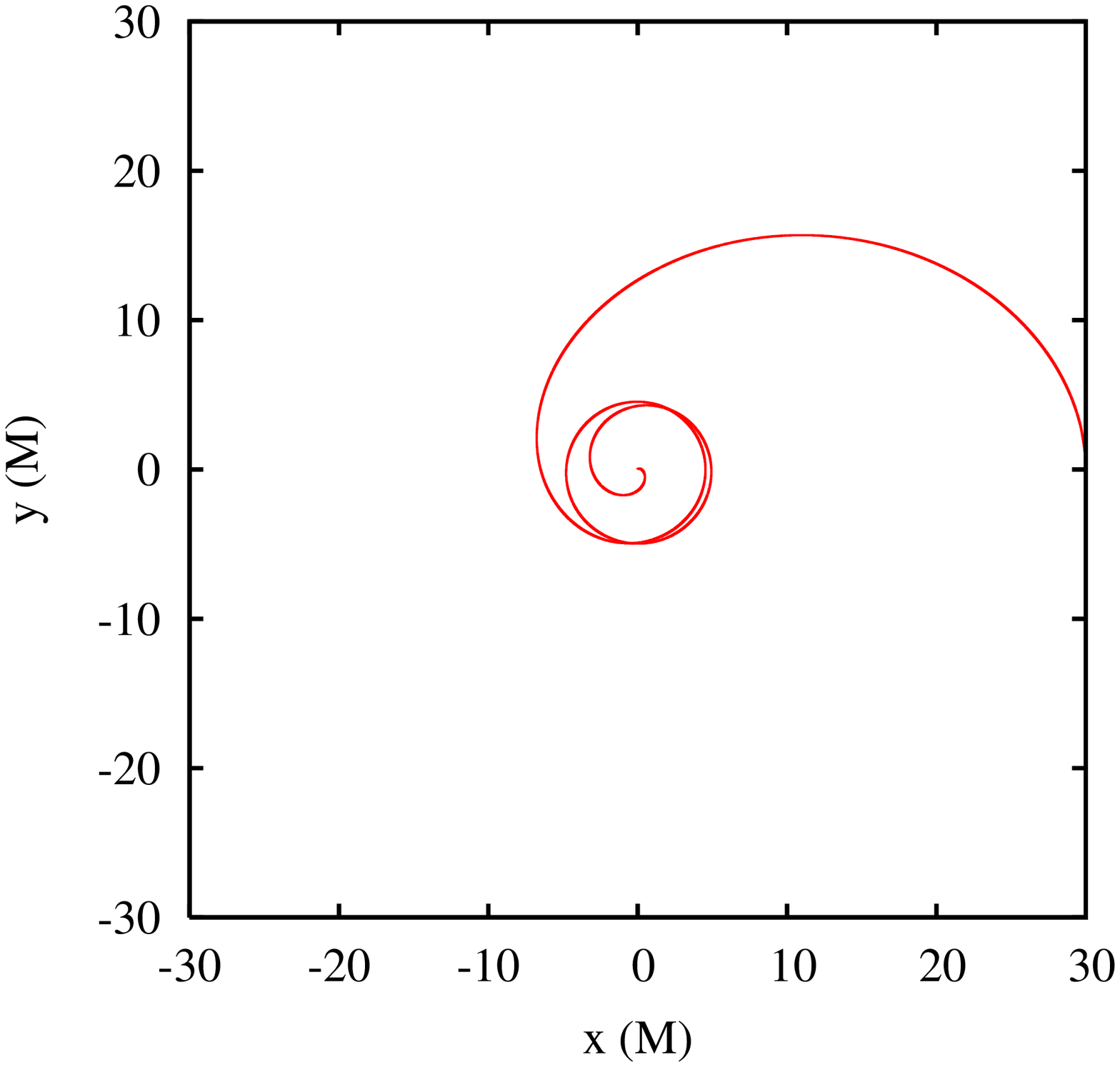}\includegraphics[width=30mm,angle=-90]{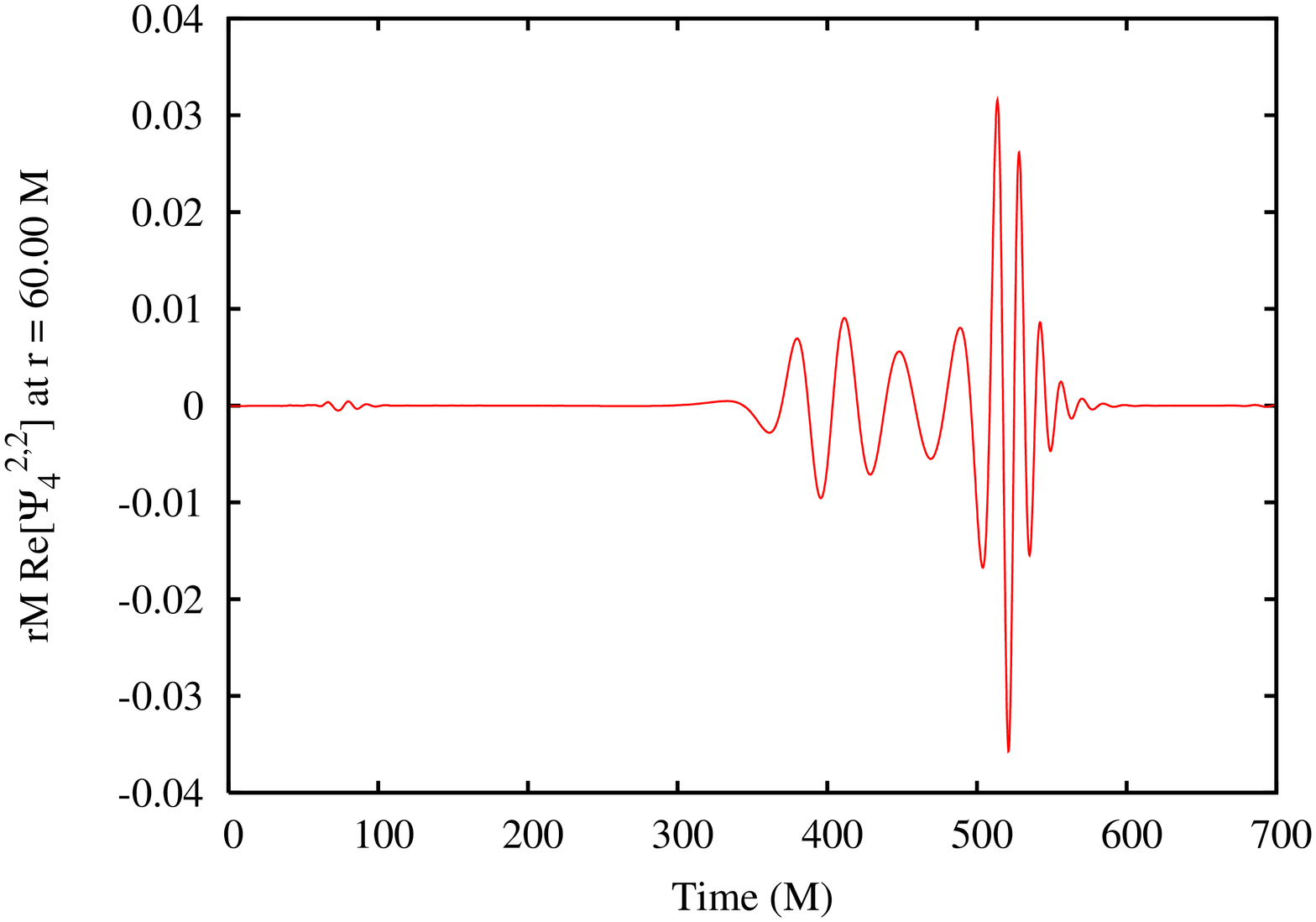}
\hfill
  \caption{
The initial apastron is
    $r_a=30M$. Left column, from the top, frame 1:  $L=4.10$ and
    the orbit rapidly moves through zoom-whirl cycles precessing by
    $>\pi$ ($q\sim 1/2$) between apastra. The orbit rapidly loses
    eccentricity but still merges with non-zero eccentricity. 
    Frame 2: $L=3.95$ and the orbit
    precesses by nearly$> 2\pi$ ($q\sim 1$) in the first radial cycle
    before merging.
    Frame 3:
    $L=3.915$ and the orbit is very nearly homoclinic, precessing by
    $\sim 4\pi$ ($q\sim 2$)  around the unstable circular orbit before 
    radiative losses cause merger. 
    Right Column:  The gravitational waveforms versus simulation time were extracted
    at a
    finite radius of $60M$. 
}
\label{NR}
\end{figure}

The runs in Fig.\ \ref{NR} were simulated using the Georgia Tech 
\texttt{MayaKranc} code that uses the same computational infrastructure
and methodology as in previous studies
\cite{2008arXiv0802.2520W,2007ApJ...661..430H,2007CQGra..24...33H,2009arXiv0905.3914H,2009PhRvL.102d1101H}, 
namely a Baumgarte-Shapiro-Shibata-Nakamura code with moving puncture gauge conditions \cite{campanelli2006,baker2005}
using the \texttt{Kranc} code generator~\cite{Husa:2004ip}.  
The black-hole encounters are initiated with Bowen-York initial 
data~\cite{Bowen:1980yu}.  The black holes are located on the 
$x$-axis: bh$_\pm$ are located at $x_+ = 7.5 M$ and $x_- = -22.5 M$ 
where $m_+ =3 M/4$ 
and $m_-= M/4$.  The spatial finite differencing is sixth order.  We 
used eleven levels of refinement with Carpet~\cite{Schnetter-etal-03b}, 
a mesh refinement package in Cactus~\cite{cactus_web}.  The finest 
resolution is $M/143$ and the outer boundary is at $287M$.  The total 
initial orbital angular momentum, is varied between the 
values of $3.9$ and $4.1$. The range is shifted from the conservative
PN value as expected, yet zoom-whirl behavior was comfortably found
given that initial prediction.

Fig.\ \ref{NR} shows three different orbits and their waveforms
corresponding to different initial
values of $L$. The largest value of $L$ shows zoom-whirl behavior
characterized by
$q\sim 1/2$ before merging. The middle value of $L$ has a $q\sim 1$ 
while the smallest value of $L$ we show is very close to the
separatrix and whirls nearly twice. 
We caution that we are only loosely reading off
these values since $q$ is changing so rapidly during inspiral.

As expected, the pairs merge by rolling over the top of the
effective potential (Fig.\ \ref{Vm.25Sm8}).
We emphasize that 
the nearly circular pattern of 
the whirl phase is not equivalent to full circularization of the
orbit. In other words, quite importantly, the orbits shown in
Fig.\ \ref{NR} do not merge through the ISCO but rather roll over the
top of the potential, merging through a whirl. To compare with the
language in a previous paper \cite{Hinder:2007qu}, some of
those eccentric orbits were circular because they merged
through a whirl phase and not because they merged through the ISCO.
This could be relevant for initial conditions for the ringdown phase.

The waveforms show distinctly quiet phases during the highly
elliptical zooms followed by louder glitches during the nearly
circular whirls. The distinctive spikes of zoom-whirls are directly
related to their rational number $q$. This suggests that zoom-whirl
orbits could be dug out of the data using algorithms suited for burst
searches, perhaps in conjunction with more targeted template
searches. 

\begin{figure}[t]
\centering
\includegraphics[width=50mm,angle=-90]{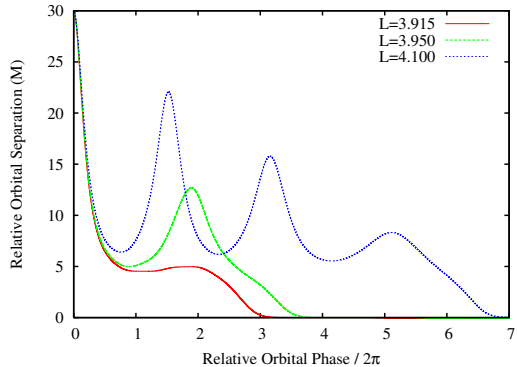}
\hfill
\caption{The relative orbital separation in units of M plotted
  versus the orbital phase for the series of $L$ in Fig.\ \ref{NR}.
}
\label{relsep}
\end{figure}

Fig.\ \ref{relsep} plots the relative orbital separation 
of the black holes versus phase for our three
cases of
$L$.  The plot illustrates the rapid merger
of the
black holes as $L$ decreases. The figure also
shows that
inspiral ends and plunge begins in the simulations for 
radii $\sim 5M$.
In line with the numerical results, 
the effective potential picture in Fig. \ref{Vm.25Sm8}
predicts whirls
around periastron, $\sim 5M$,
which is much less than the PN predicted value of the ISCO,
$r_{ISCO}\sim 8.8M$, for a mass ratio of 1/3 and these spins. 
The figure therefore confirms that the zoom-whirl pair merge near a
whirl phase and not near the ISCO.
The
top line, corresponding to $L=4.1$, shows the three zooms before
merger that we see in frame 1 of Fig.\ \ref{NR}.   The bottom line,
corresponding to $L=3.915$ and frame 3 of Fig.\ \ref{NR} is almost
homoclinic with no zooms, precessing around the unstable circular
orbit before merging.
%

In general, we conclude from the PN approximation that
zoom-whirl behavior has better odds at surviving
dissipation and leaving a mark in the gravitational waveform for (1)
more disparate values of $m_2/m_1$ because of the slower dissipation
time, 
(2) larger spin magnitudes, all else being equal, because of
the greater frame dragging effect, (3)
larger initial apastra because of the longer orbital lifetime. 
There is less zoom-whirl behavior in equal-mass nonspinning pairs, hence the
importance of sticking closer to the separatrices 
\cite{pretorius2007}. Yet more generally, zoom-whirl orbits
ought to extend down below the separatrices.
In the future, we intend to
extend this work by pushing the numerical simulations to
large mass ratio and extracting a measure of $\dot q$.

For any zoom-whirl pair, the glitchy waveforms followed by longer
quiescent phases are highly distinctive and beg for tailored search
algorithms.
Furthermore, the loud glitches could 
nudge these highly
precessing orbits into an optimistic position for early direct detection.

{\bf Acknowledgements:} The authors wish to acknowledge NSF grants
AST-0908365, PHY-0925345,  PHY-0653303 and TG-PHY060013N that supported this work.
We also thank Ian Hinder, Frank Herrmann,Tanja Bode and Pablo Laguna for
contributions to the \texttt{MayaKranc} code.

\bibliographystyle{aip}
\bibliography{gr,nr}

\end{document}